\documentclass[aps,pre,showpacs,twocolumn,floatfix,longbibliography]{revtex4-1}
\usepackage{amsfonts}
\usepackage{amsmath}
\usepackage{amsthm}
\usepackage{graphics}
\usepackage{graphicx}
\usepackage{subfigure}
\usepackage{amssymb}
\usepackage{color}
\usepackage[T1]{fontenc}
\usepackage{amsmath,amssymb,color,epsfig,latexsym,natbib,graphicx,dcolumn}

\newcommand{\bomega}{\mbox{\boldmath{$\omega$}}}

\newcommand{\R}{\mathbb{R}}

\newcommand{\mbf}[1]{{\mathbf{#1}}}

\newcommand{\bu}{{\bf u}}

\newcommand{\dt}{\Delta t}
\newcommand{\nkp}{\nu k^{2p}}
\newcommand{\mN}{\mathcal{N}}

\begin{document}
\title{Turbulent cascade, bottleneck and thermalized spectrum in hyperviscous flows} 

\author{Rahul Agrawal}
\affiliation{Department of Mechanical Engineering, Indian Institute of Technology Bombay, Maharashtra, India 400076}

\author{Alexandros Alexakis}

\author{Marc E. Brachet}
\affiliation{Laboratoire de Physique de l'\'Ecole Normale Sup\'erieure, CNRS,
    PSL Research University, Sorbonne Universit\'e, Universit\'e de Paris,
    F-75005 Paris, France}
    
\author{Laurette S. Tuckerman}
\affiliation{Physique et M\'ecanique des Milieux 
H\'et\'erog\`enes (PMMH), ESPCI Paris, 
CNRS, PSL Research University, Sorbonne Universit\'e, Universit\'e de Paris, F-75005 Paris, France}

\date{\today}


\begin{abstract}
In many simulations of turbulent flows the viscous forces $\nu\nabla^2 \bu$ are replaced by a hyper-viscous term $-\nu_p(-\nabla^2)^{p}\bu$ in order to suppress the effect of viscosity at the large scales. In this work we examine the effect of hyper-viscosity on decaying turbulence for values of $p$ ranging from $p=1$ (regular viscosity) up to $p=100$. Our study is based on direct numerical simulations of the Taylor-Green vortex for resolutions from $512^3$ to $2048^3$. 
Our results demonstrate that the evolution of the total energy $E$ and   the energy dissipation $\epsilon$ remain almost unaffected by the order of the hyper-viscosity used. 
However,  as the order of the hyper-viscosity is increased, the energy spectrum develops a more pronounced  bottleneck that contaminates the inertial range. At the largest values of $p$ examined, the spectrum at the bottleneck range has a positive power-law 
behavior $E(k)\propto k^\alpha$ with the power-law exponent $\alpha$ approaching the  value obtained in flows at thermal equilibrium $\alpha=2$. 
This agrees with the prediction of Frisch et al. [Phys. Rev. Lett. 101, 144501 (2008)] who suggested that at high values of $p$, the flow should behave like the truncated Euler equations (TEE). Nonetheless, despite the thermalization of the spectrum, the flow retains a finite dissipation rate up to the examined order, which disagrees with the predictions of the TEE system implying suppression of energy dissipation. We reconcile the two apparently contradictory results, predicting the value of $p$ for which the hyper-viscous Navier-Stokes goes over to the TEE system and we discuss why thermalization appears at smaller values of $p$.
\end {abstract}
\maketitle
\section{Introduction} \label{Sec:Intro}

Most planetary and astrophysical flows are highly turbulent. As a result, for a wide range of scales, the viscosity has no direct effect on the flow and so the flow evolves as if it were inviscid. Nonetheless viscosity cannot be neglected because it acts effectively at the smallest scales, converting the coherent energy of the flow into heat. It is thus essential in numerical simulations to resolve all scales: from the large scales, where energy is injected and which follow inviscid dynamics, to the smallest scales, where dissipation takes place. However, even with today's computational power it is still impossible to achieve a resolution which is sufficient to simulate most atmospheric flows. Various methods have therefore been devised to model the small-scale dissipation in order to follow the inviscid dynamics of the large-scale flows while correctly capturing the dissipation rate at small scales. A popular model for spectral codes is the use of hyper-viscosity, meaning that the Laplacian of the standard viscous term is replaced by a higher power of the Laplacian. In this way the portion of the spectral resolution devoted to simulating the viscous wavenumbers is reduced, leaving a larger range of wavenumbers that evolve almost inviscidly. Indeed, hyperviscosity models have been shown to reproduce  the turbulent evolution  of the large scales, manifesting the Kolmogorov energy spectrum 
\begin{equation} E(k) = C \epsilon^{2/3}k^{-5/3}, \end{equation} 
where $E(k)$ is the energy spectrum, $k$ is the wavenumber, $\epsilon$ 
is the energy dissipation rate and $C\simeq1.58$ is the Kolmogorov constant.

However, the statistics of turbulent flows are not unaffected by this change of the dissipation term and various studies have been devoted to analyzing these undesirable effects of hyper-viscosity 
\cite{borue1995forced,
borue1996numerical, 
l1998universal,
haugen2004inertial, 
lamorgese2005direct,
frisch2008hyperviscosity, 
spyksma2012quantifying}.
In particular, at small scales close to the dissipation range, hyper-viscosity is known to produce an aggravated {\it bottleneck} effect. 
The bottleneck effect is an accumulation of the cascading energy 
at wavenumbers just below the dissipation range, leading to a change in the power-law behavior of the energy spectrum. The bottleneck effect exists even for ordinary viscosity and has been the subject of many studies in turbulence
\citep{
falkovich1994bottleneck, 
lohse1995bottleneck, 
martinez1997energy, 
donzis2010bottleneck, 
kuchler2019experimental}.
With the use of hyper-viscosity, the bottleneck becomes more pronounced, even leading to
a non-monotonic behavior of the energy spectrum. In fact it has been conjectured by Frisch et al.~\cite{frisch2008hyperviscosity} that for sufficiently high order of hyper-viscosity, 
the bottleneck will take the form of a {\it thermalized} absolute equilibrium state discussed by Kraichnan~\citep{kraichnan1973helical} in which energy is equally distributed among all Fourier modes, leading to an energy spectrum proportional to $k^2$. This thermalized spectrum is realized in the truncated Euler equations, for which the Euler equations are solved in Fourier space while keeping only a finite number of Fourier modes.
This system conserves exactly the inviscid quadratic invariants of the system.
The argument of \cite{frisch2008hyperviscosity} for the appearance of the thermalized 
energy spectrum in hyper-viscous flows is that as the order of the hyper-viscosity 
is increased to very large values, it suppresses all energy above a wavenumber 
$k_{_G}$ (determined by the value of the hyper-viscous coefficient) while leaving 
unaffected all wave numbers below. It thus acts as a Galerkin truncation, similarly to the truncated Euler equations. This result was verified with the use of the EDQNM approximation and for the one-dimensional Burgers equation.
However, it has not been verified for the three-dimensional hyper-viscous Navier-Stokes equations. 

In this work, we explore further the effect of hyper-viscosity carrying out high-resolution Direct Numerical Simulations (DNS) of the decay of a Taylor-Green vortex. Taking advantage of the symmetries of the Taylor-Green flow \cite{brachet2013ideal} and using a
slaved time-stepping method \cite{frisch1986viscoelastic}
we have been able to perform many simulations at different orders of hyper-viscosity, reaching values that are sufficiently high 
to test the thermalisation conjecture.

\section{Definition of the system \label{Sec:Theo} }

\subsection{Basic definitions}

We consider the 3D hyper-viscous incompressible Navier-Stokes equations 
that control the evolution of the velocity field
$\mbf{u}(x,y,z,t) \in \R^3$ defined in $(x,y,z) \in [0,2\pi L]^3$ and in a time interval $t \in [0,T)$:
\begin{eqnarray}
\label{eq:NS}
 \frac{\partial \mbf{u}}{\partial t} + \mbf{u} \cdot \nabla \mbf{u} = - \nabla P - \nu_p (-\nabla^2)^{p}\mbf{u}, 
\end{eqnarray}
where incompressibility $\nabla \cdot \mbf{u} = 0$ is assumed, $P$ is the pressure, $p$ is the order of hyper-viscosity, and $\nu_p$ a hyper-viscosity coefficient. 
The periodicity of ${\bf u}$ allows us to
use the (standard) Fourier representation
\begin{eqnarray}
 \widehat{\mbf{u}}(\mbf{k},t) &=& \frac{1}{(2 \pi L)^3} \int_D {\bf u}({\bf x},t) \exp(-i \mbf{k}{\bf x}) d^3x \\
{\mbf{u}}(\mbf{x},t) &=& \sum\limits_{\mbf{k}\in \mathbb{Z}^3}  \widehat{\mbf{u}}(\mbf{k},t) \exp(i \mbf{k}{\bf x})  \label{eq:Four2},
\end{eqnarray}
The kinetic energy spectrum $E(k,t)$ is defined as the sum over spherical shells
\begin{equation}
\label{eq:spectrum}
E(k,t) = \frac{1}{2} {\displaystyle \sum_{\underset {k-1/2  < |\mbf{k}| < k+1/2}{\mbf{k} \in \mathbb{Z}^3} }} |\widehat{\mbf{u}}({\bf k},t)|^2,
\end{equation}
and the total energy is
\begin{equation}
E =\frac{1}{2 (2 \pi L)^3} \int_D {\left|{\bf u}({\bf x},t)\right|^2} d^3x = \frac{1}{2} {\displaystyle \sum_{\mbf{k} \in \mathbb{Z}^3}} |\widehat{\mbf{u}}({\bf k},t)|^2 ,
\end{equation}
The dissipation rate of energy is given by
\begin{equation}
\epsilon=\nu_p {\displaystyle \sum_{\mbf{k} \in \mathbb{Z}^3}} k^{2p}|\widehat{\mbf{u}}({\bf k},t)|^2 .
\label{eq:diss}\end{equation}
As $p$ increases, the dissipation is concentrated at increasingly large wavenumbers.
We calculate the dissipation rate as the finite difference $-dE/dt$, since summing the expression \eqref{eq:diss} would multiply the small error in $E(k)$ by a large factor $\nkp$.

\subsection{Taylor-Green vortex}

The initial condition we consider is the Taylor-Green (TG) vortex
\cite{TG1937}, which is given by 
\begin{equation}
{\bf u}^{\mathrm{TG}}=U \left[
\begin{array}{c}
+\sin(x/L) \cos(y/L)\cos(z/L) \\
-\cos(x/L) \sin(y/L)\cos(z/L) \\
0
\end{array}\right] \label{eq:TYGID}
\end{equation}
so that the total energy  is given by $E=U^2/8$. We non-dimensionalize by $L$ and $U$, setting these to 1.
Time is scaled by the advective time ($L/U = 1$).

The TG vortex is closely related to the von K\`arm\`an (VK) swirling flow that has been the subject of many experimental studies \cite{DouadyPRL,Fauve1993,Maurer1994}.
The VK flow and the TG vortex have the same basic geometry: both consist of a shear layer between two counter-rotating circulation cells. The TG vortex, however, is periodic with impermeable free-slip boundaries (present as mirror symmetries) while the experimental flow takes place between two counter-rotating coaxial impellers and is confined inside a cylindrical container. The TG vortex also obeys a number of additional rotational symmetries.

The symmetries of the TG initial conditions \eqref{eq:TYGID} are 
preserved by the time evolution. 
These are, first, rotations by $\pi$ around the axes $x=z=\pi/2$ and $y=z=\pi/2$,
and by $\pi/2$ around the axis $x=y=\pi/2$.
A second set of symmetries corresponds to planes of mirror symmetry:
$x=0,\pi$, $y=0,\pi L$ and $z=0,\pi L$. On the symmetry planes, the
velocity ${\bf u}^{\mathrm{TG}}$ and the vorticity
${\bomega^\mathrm{TG}}={\bf\nabla} \times {{\bf u}^{\mathrm{TG}}}$ are
(respectively) parallel and perpendicular to these planes that form
the sides of the so-called impermeable box which confines the flow.
It is demonstrated in \cite{BRACHET:1983p4817} that these symmetries
imply that the Fourier expansion coefficients $\widehat{\mbf{u}}(m,n,p,t)$ of the velocity field
in \eqref{eq:Four2} vanish unless
$m,n,p$ are either all even or all odd integers. This can be used
to reduce memory storage and speed up computations \cite{PhysRevE.78.066401,PLBMR2010}
by a factor of 8. If this symmetry is not imposed, round-off errors can break the symmetries
as the flow evolves.
However, this bifurcation occurs significantly later than the times
considered in the present study.

To simulate the evolution of the Taylor-Green flow we used the TYGRS (TaYlor-GReen Symmetric), a pseudospectral parallel code which enforces the
symmetries of the TG vortex in 3D hydrodynamics within the periodic
cube of length $2\pi$. Details of the code can be found in \cite{brachet2013ideal}.

 \subsection{Choice of parameters}

As the value of $p$ is varied, the value of $\nu_p$ must be adjusted accordingly. To attain high Reynolds numbers, $\nu_p$ should be as small as possible, subject to the constraint that the simulation be well resolved.
To insure adequate resolution, 
we measured the energy spectrum $E(k)$ at the time of maximum energy dissipation and verified that at large $k$
it follows an exponential law $E(k) \propto e^{-  k/k_d}$ with $k_d$ the dissipation wavenumber such that $k_{max}/k_d\ge 2$. Here $k_{max}$ is the maximum wavenumber given by $k_{max}=N/3$ due to de-aliasing. The exponential law implies that the grid size is smaller than the hyper-viscous Kolmogorov lengthscale
$\eta_p = (\nu_p^{3}/\epsilon)^{1/(6p-2)}$ (where $\epsilon \propto U^3/L$) i.e. that $k_{max}\, \eta_p < 1$.
This in turn implies that for a fixed grid size, $\nu_p$ should be chosen to have an exponential dependence on $p$ given by
$ \nu_p \propto U L^{-1/3} k_{max}^{2p-2/3} $.
If we define the Reynolds number $Re_p$ to be inversely proportional to the hyper-viscosity 
as $Re_p = UL^{2p-1}/\nu_p$ then the value that can be achieved for a given resolution is 
\begin{equation} 
Re_p \propto (k_{max}L)^{2p-2/3}.
\label{eq:Rep}
\end{equation}

The high value of $p$
imposes additional demands on the time-integration scheme.
Because of this, we used a modified exponential method, also called the {\it slaved}
method \cite{frisch1986viscoelastic}, which is described in detail in Appendix \ref{appendix}.

\begin{figure}[ht]
 \begin{center}
 \includegraphics[width=1\columnwidth]{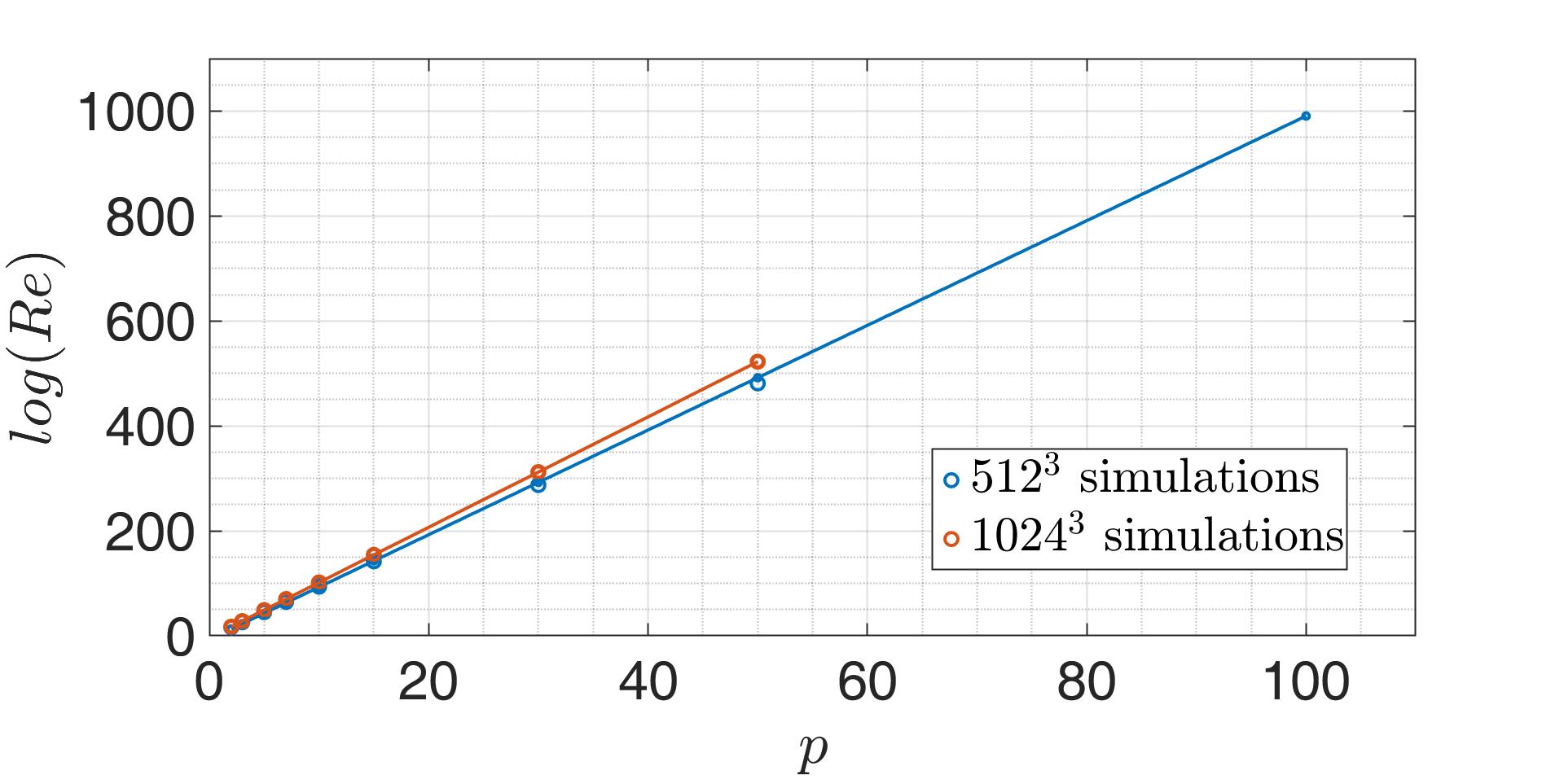}
 \caption{Reynolds number used in the different runs as a function of the order $p$ of hyper-viscosity.  }
 \label{fig:Rep}
 \end{center}
 \end{figure}
 In the present study, we carry out two series of simulations, 
 one with resolution $N=512$, and another with resolution $N=1024$, in which 
 we varied the value of $p$ from 1 to 100 or
 from 1 to 50, respectively.
 The scaling \eqref{eq:Rep} of the parameters for our runs
 is shown in figure \ref{fig:Rep} for the two resolutions.
 In addition, we performed a simulation with $p=1$ 
 at $N=2048$, which serves as a baseline case with which to compare our hyper-viscous runs.
 

\section{Results\label{Sec:Results}} 

\subsection{ Global Dynamics  }

\begin{figure}[ht]
 \begin{center}
 \includegraphics[width=1\columnwidth]{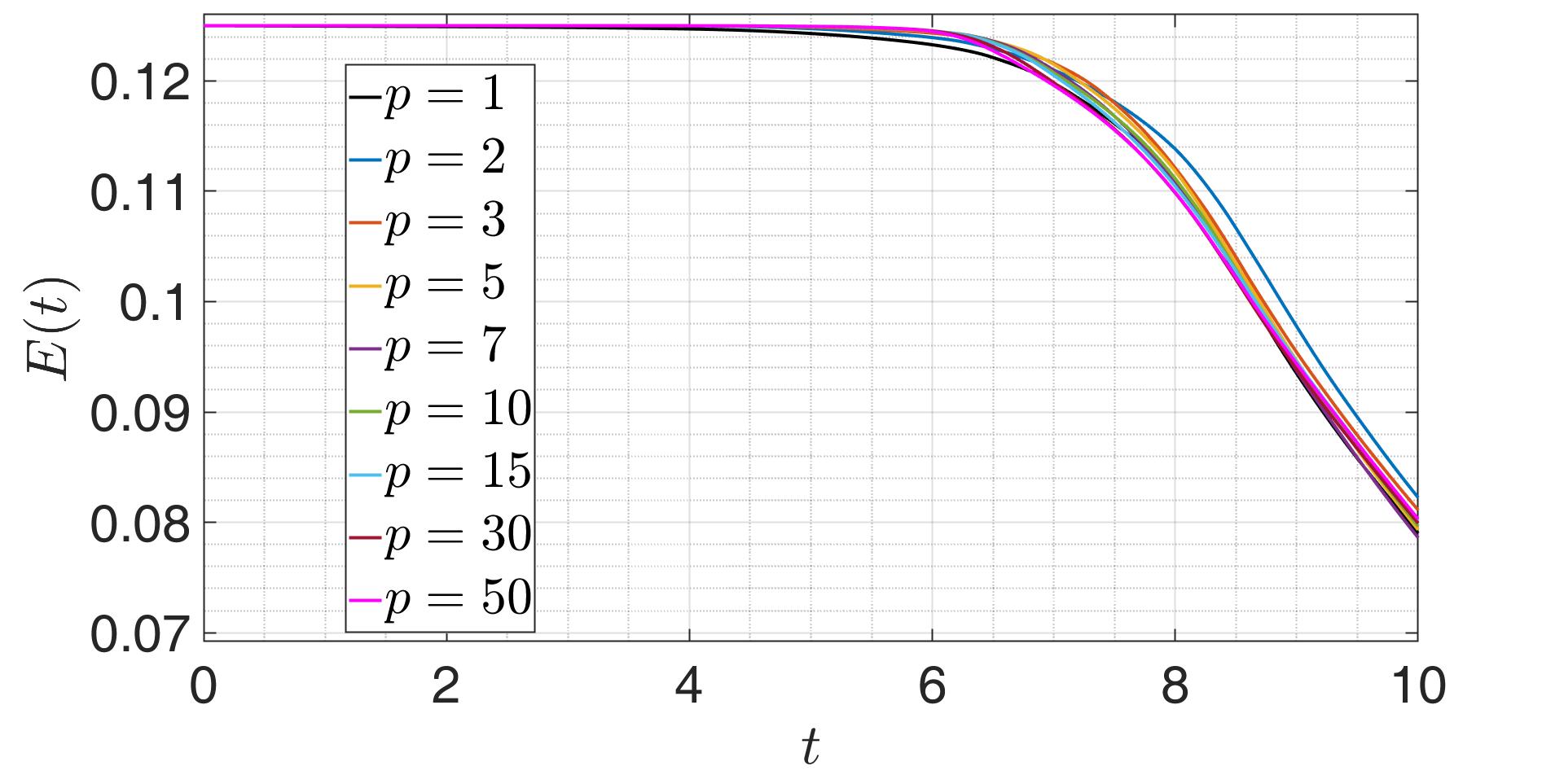}
 \includegraphics[width=1\columnwidth]{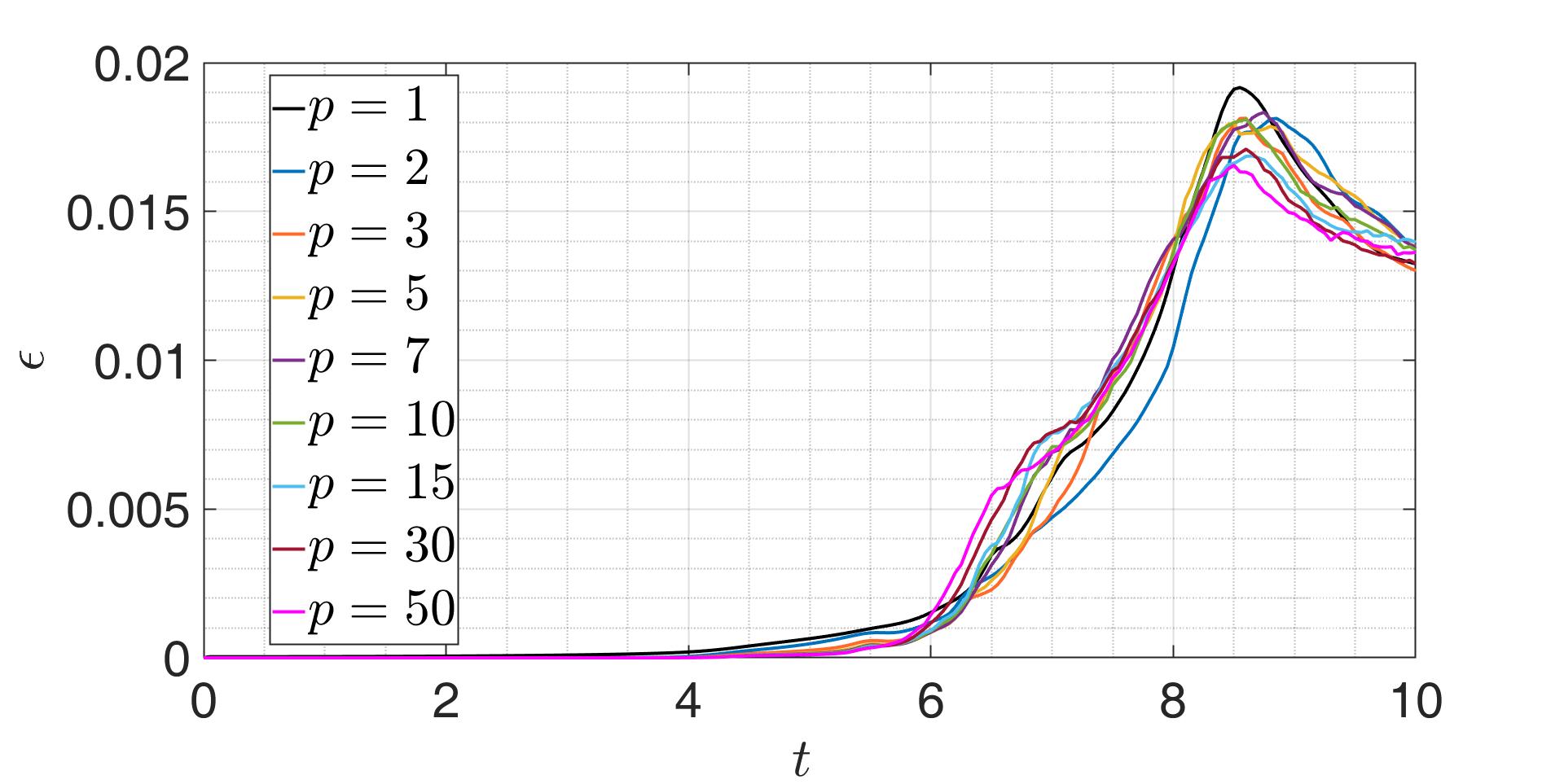}
 \caption{Top: Time evolution of the total energy for $1024^3$ resolution runs 
 for different $p$ values compared to the $p=1$, $N=2048$ run. 
Bottom: mean dissipation rate as a function of time for the same runs.  }
 \label{fig:EnergyDiss}
 \end{center}
 \end{figure}

The top panel of figure \ref{fig:EnergyDiss} shows the evolution of the energy
as a function of time from the simulations at resolution $N=1024$.
The bottom panel shows the energy dissipation rate.
In both panels, the results are compared with the results from the simulation
with ordinary viscosity and the higher numerical resolution $N=2048$.

Surprisingly, even for the largest values of $p$ used, both the energy and the energy dissipation rate are very close to those of the high-resolution run with ordinary viscosity, with the peak dissipation occurring at $t\simeq 8.5$. Thus, despite the very different mechanisms used to dissipate the energy, the global dynamics of the system have not been altered. This
indicates that at this resolution, the rate that energy is dissipated is controlled by the large scale dynamics and the energy cascade and not the exact dissipation mechanism.
 
\subsection{ Energy Spectra  }

\begin{figure}[t]
 \begin{center}
\includegraphics[width=1\columnwidth]{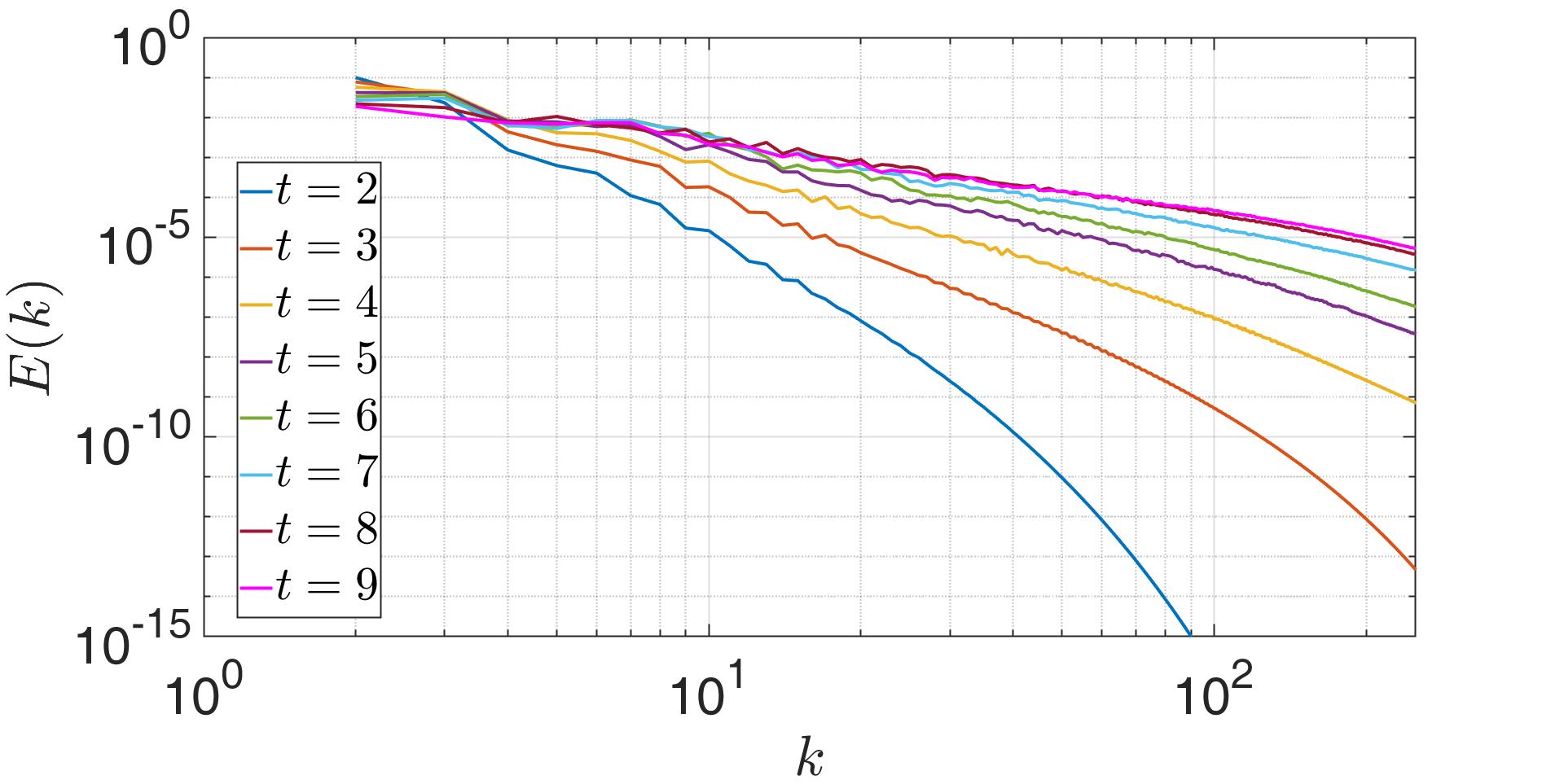}
 \includegraphics[width=1\columnwidth]{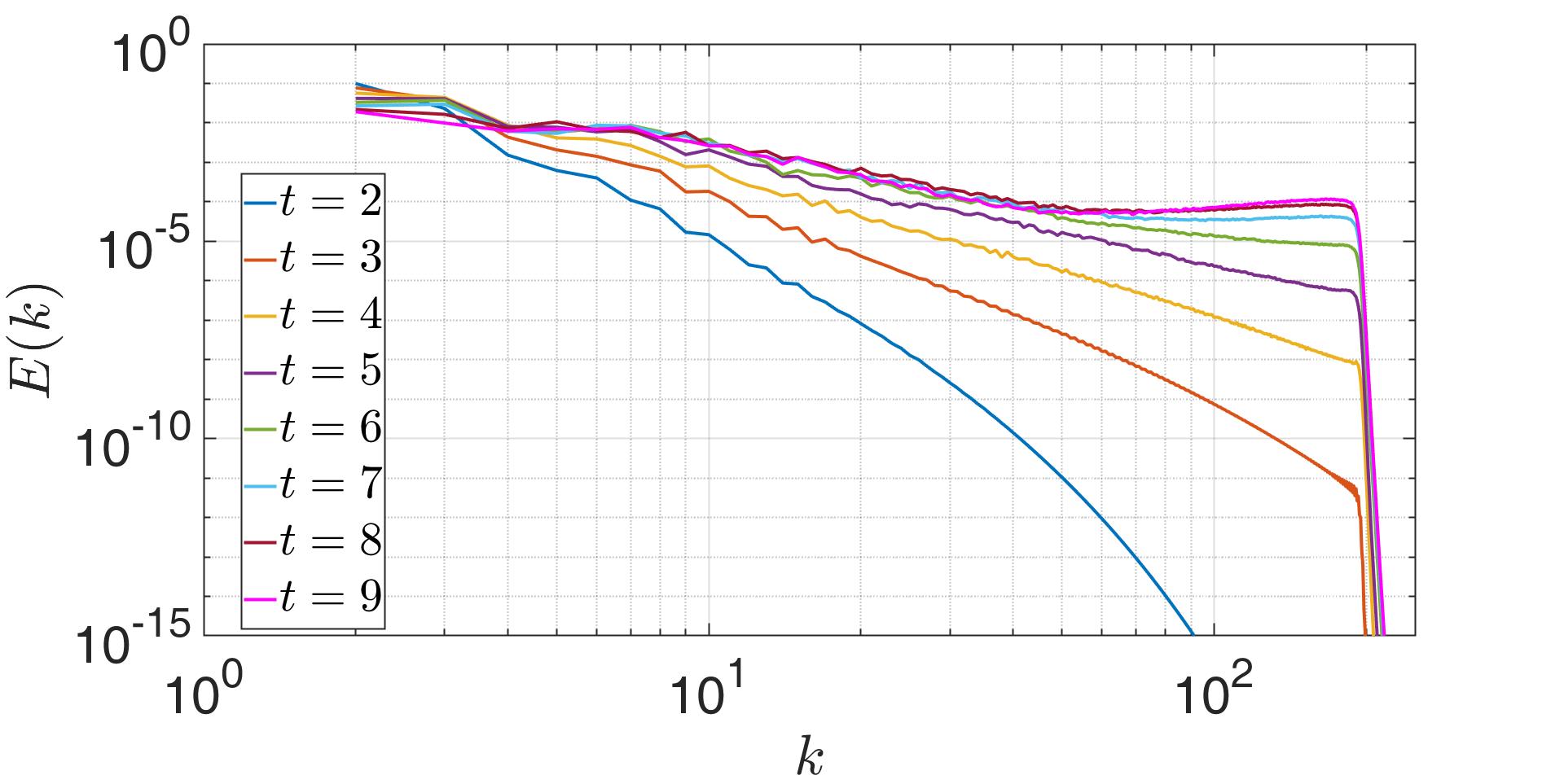}
 \caption{Energy spectra for the $p=1$, $N=2048$ 
Navier Stokes simulations (top panel) and the $p=50$, $N=1024$ simulations (bottom panel). }
 \label{fig:spectra}
 \end{center}
 \end{figure}

The global dynamics alone do not, however, guarantee that hyperviscosity correctly models the effect of turbulence at the larger scales.  In figure \ref{fig:spectra}, we show in
the top panel the energy spectra from the simulation with ordinary viscosity and resolution $2048^3$ at eight different times,
and in the bottom panel the energy spectra from the $p=50$ simulation at resolution $1024^3$.
The main and intentional effect of hyperviscosity is clear: 
there is a very rapid fall-off of the spectrum for $k\gtrsim 190$.
Looking at the spectrum for $k\lesssim 190$,
the two cases have very similar spectra for times smaller than $t=5$, for which dissipation effects are negligible,  At later times, however, the $p=50$ runs show an excess of energy at high wavenumbers that becomes more apparent as the peak of the energy dissipation at $t\simeq 8.5$ is approached.

\begin{figure}[t]
 \begin{center}
 \includegraphics[width=1\columnwidth]{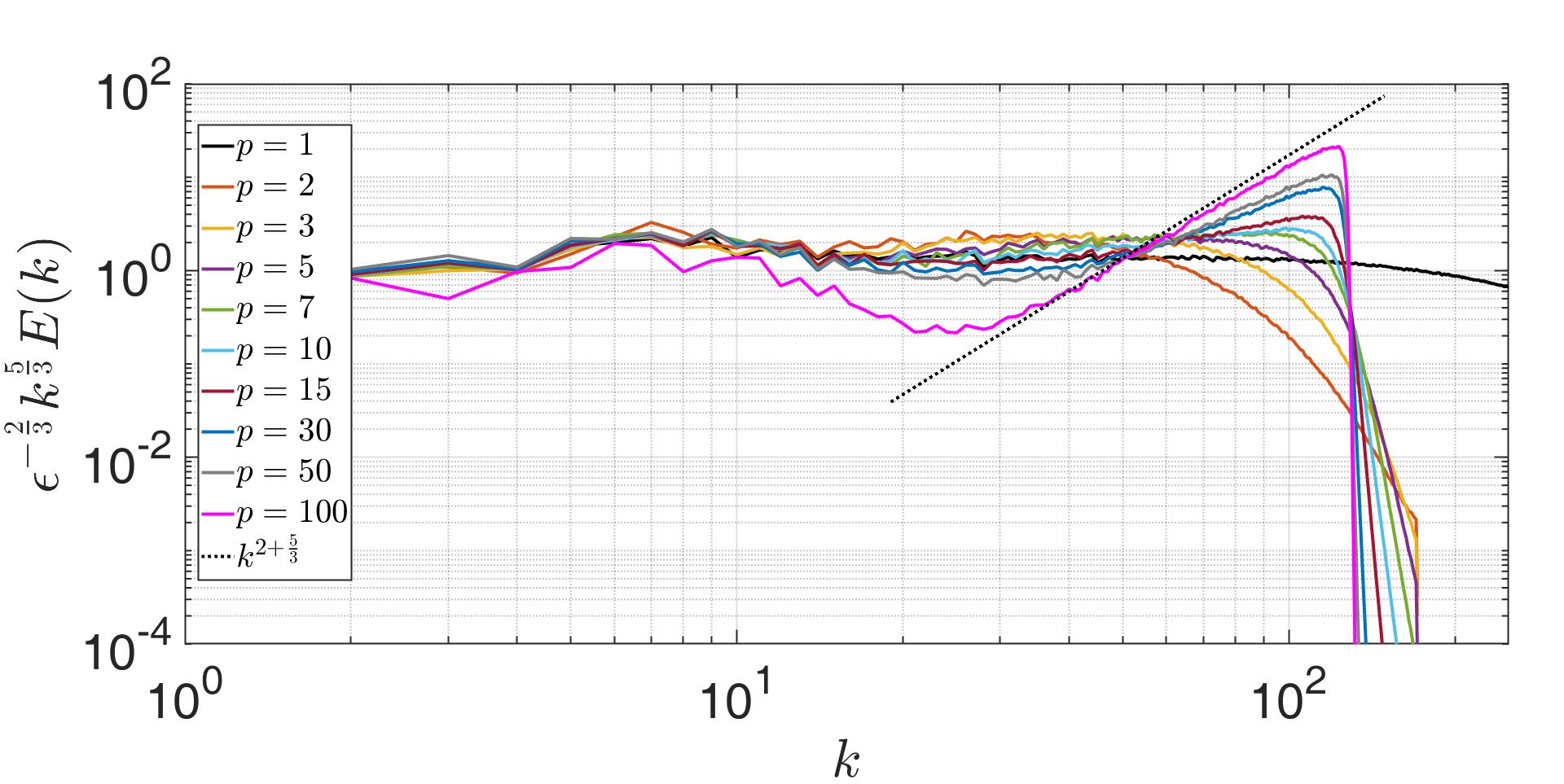}
 \includegraphics[width=1\columnwidth]{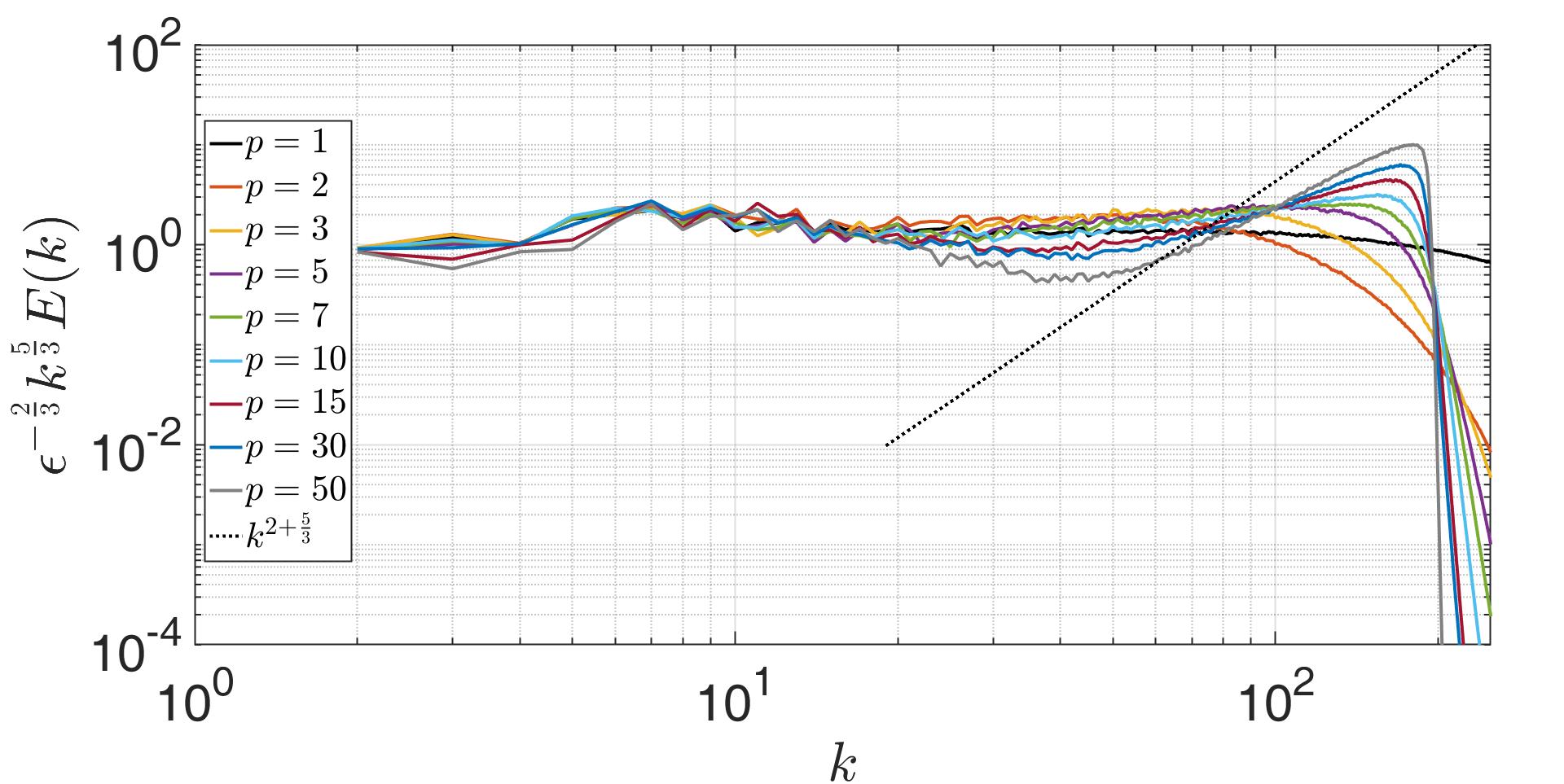}
 \caption{Top panel: Energy spectra normalized by $k^{-5/3}$ for the $512^3$
 numerical simulations (top panel) and the $1024^3$ simulations (bottom panel).}
 \label{fig:compensated_spectra}
 \end{center}
 \end{figure}

In figure \ref{fig:compensated_spectra}
we focus on the time of maximum dissipation and show the energy spectra at this instant for different values of $p$. The top panel shows the spectra from the $512^3$ resolution
numerical simulations while the bottom panel shows the results from the $1024^3$ simulations.
The spectra have been multiplied 
by $k^{5/3}$ so that a Kolmogorov spectrum would appear as flat. 
The spectra are surprisingly insensitive to the value of $p$. For all $p$, the spectra fall off for $k\gtrsim 130$ for resolution $512^3$ and for $k\gtrsim 190$ for resolution $1024^3$.
The spectra appear close to flat for wavenumbers smaller than $k=20$.
For larger wavenumbers, we observe a dip in the spectrum followed by a bottleneck, which becomes stronger as the order of the hyper-viscosity is increased.
For large values of $p$, this bottleneck takes the form of a power-law 
$E(k)\propto k^{\alpha_p}$ that increases with $p$. 

\begin{figure}[ht]
 \begin{center}
 \includegraphics[width=1\columnwidth]{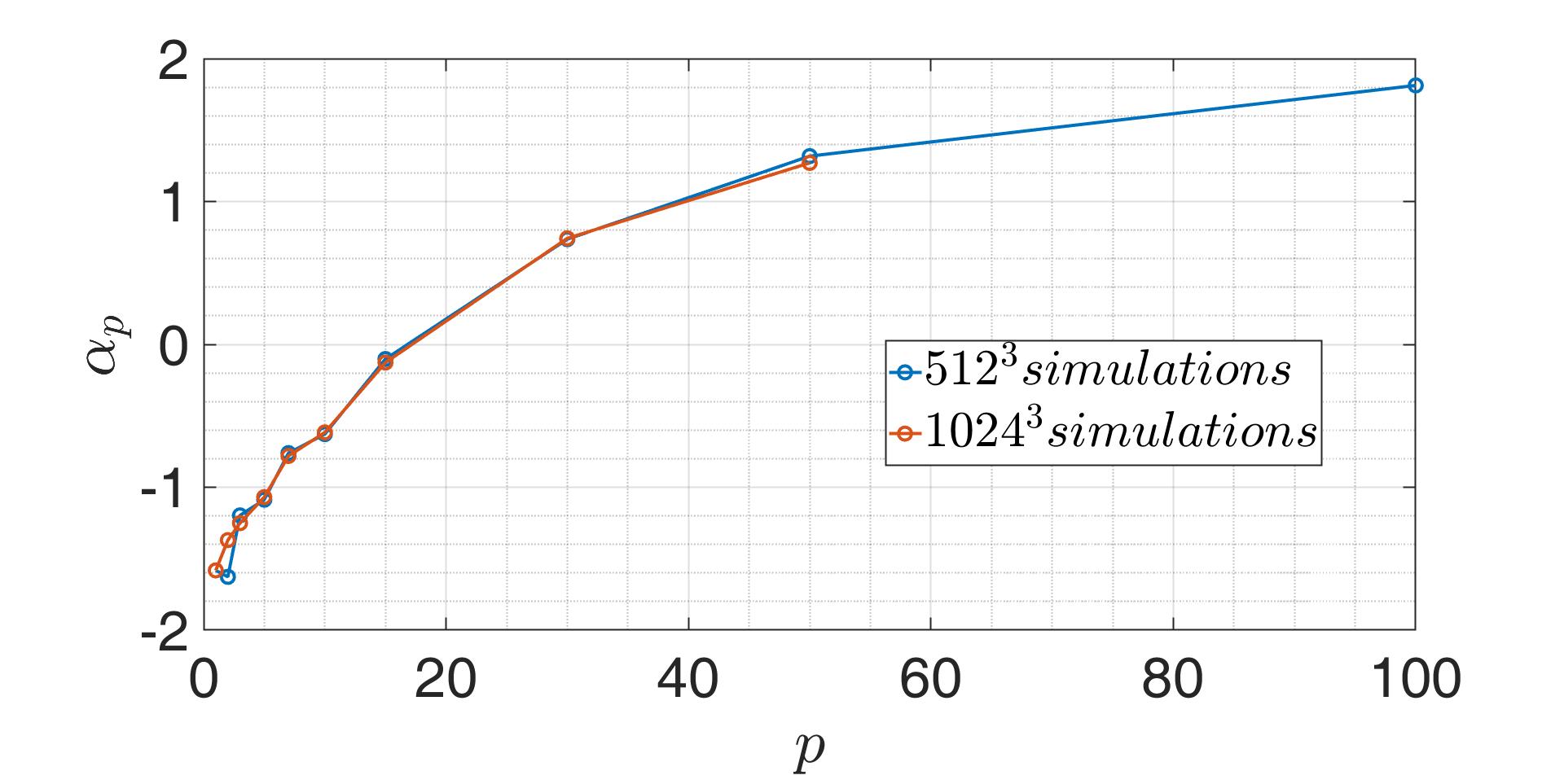}
 \caption{Measured values of $\alpha_p$ as a function of $p$ for the $512^3$ and $1024^3$ simulations. The value at $p=1$ is taken from the $2048^3$ simulations. 
 }
 \label{fig:alpha_p}
 \end{center}
 \end{figure}

The dashed line indicates 
the prediction from thermalization $E(k) \propto k^2$.
For each $p$, we obtained the value of the exponent $\alpha_p$ by 
fitting the spectrum over the increasing portion of the bottleneck, i.e. for $k$ in the range 50-70 in the $512^3$ simulations and 90-110 in the $1024^3$ simulations.
In figure \ref{fig:alpha_p} we show the value $\alpha_p$ as a function of $p$. 
We see that as $p$ becomes large, the exponent approaches 
the thermalized value $\lim_{p\to \infty} \alpha_p=2$ as predicted in \cite{frisch2008hyperviscosity}.

\subsection{ Spatial Structures }

\begin{figure*}
{\large
  a)\includegraphics[width=.45\linewidth]{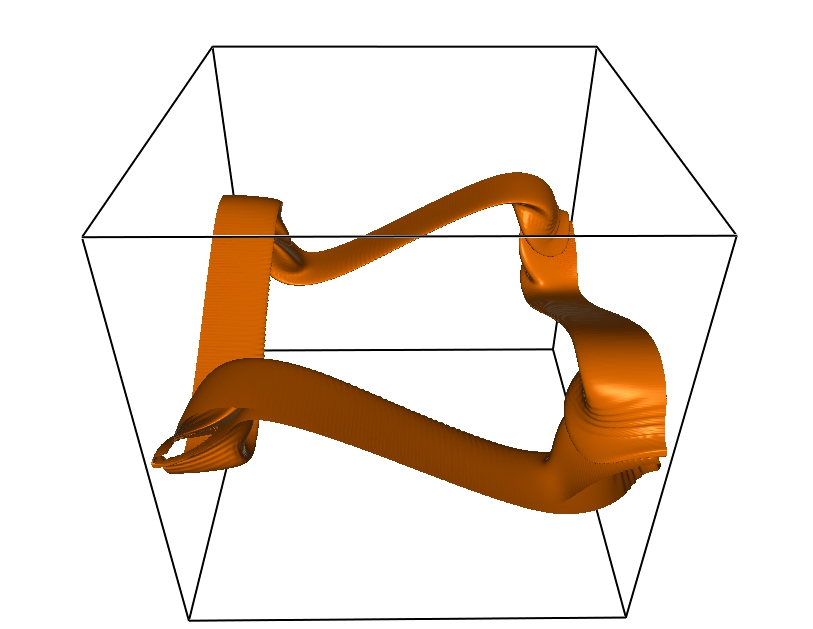}
  b)\includegraphics[width=.45\linewidth]{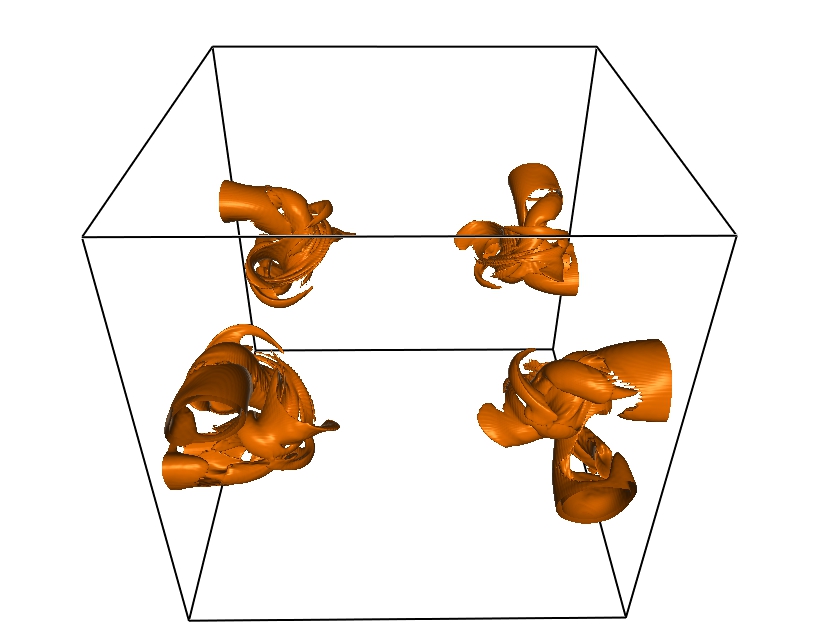}\\
  c) \includegraphics[width=.45\linewidth]{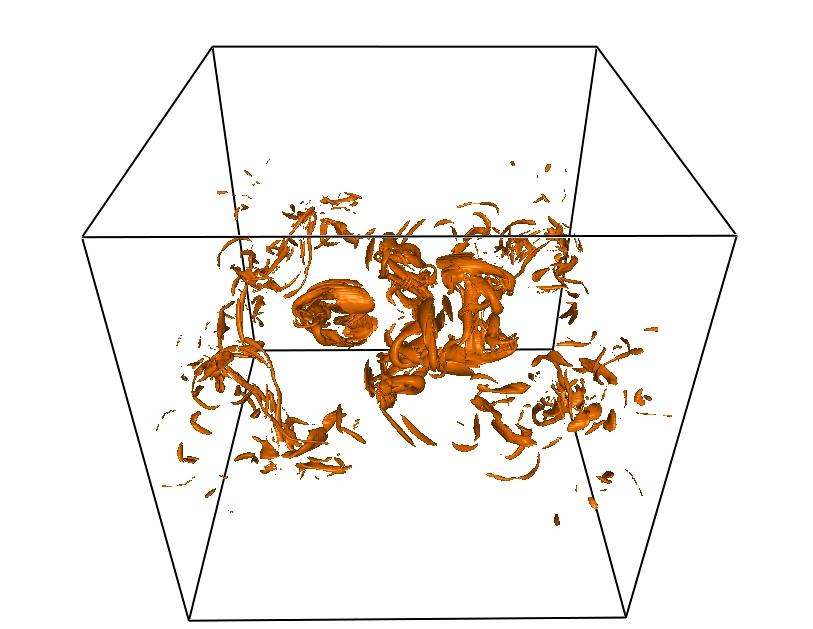}
  d)\includegraphics[width=.45\linewidth]{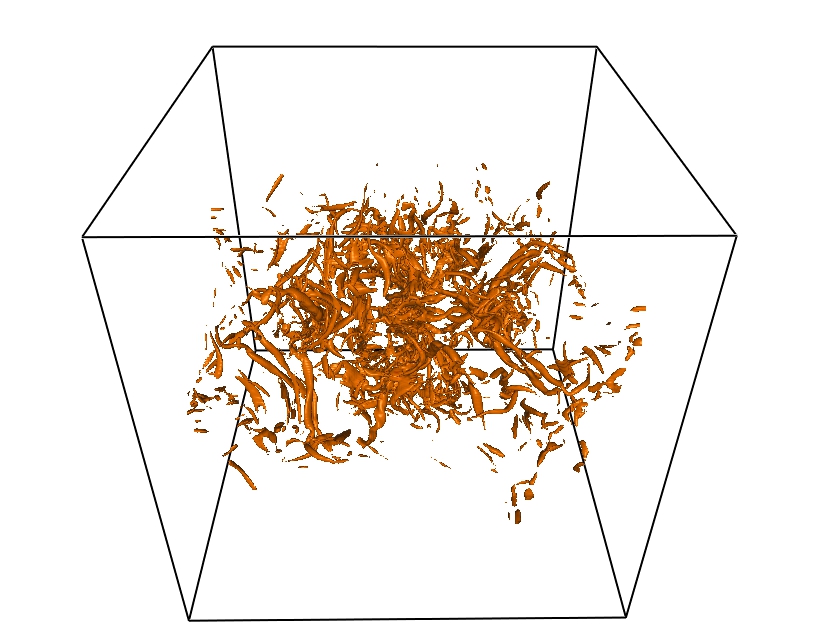}
  }
   \caption{ Isosurfaces (at $8\%$ of maximum) of square vorticity $\omega^2=(\nabla \times {\bf v})^2$ at resolution $512^3$ in the impermeable box, with $p=1$ at $t=4$, 6, 8, 10.}
 \label{fig:viz512_p_1}
\end{figure*}

\begin{figure*}
{\large
a)\includegraphics[width=.45\linewidth]{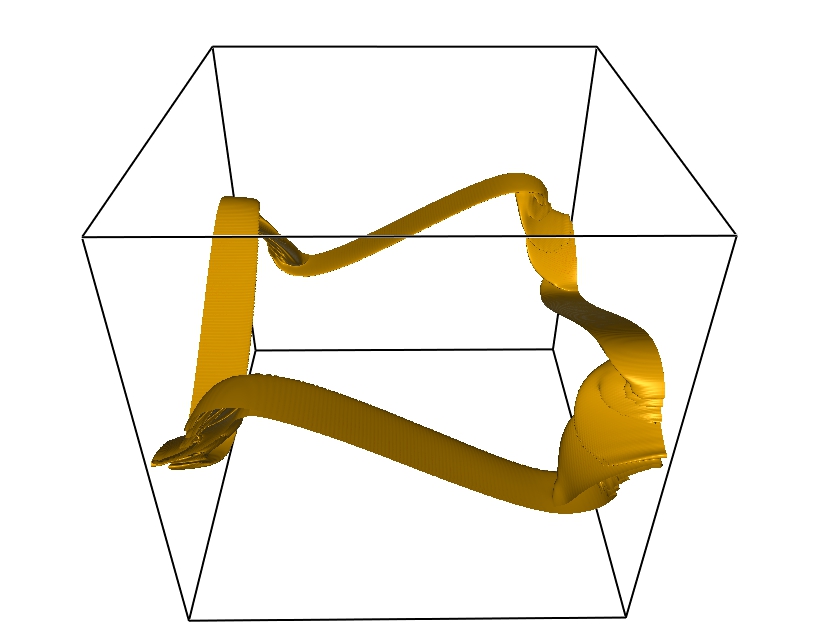}
b)\includegraphics[width=.45\linewidth]{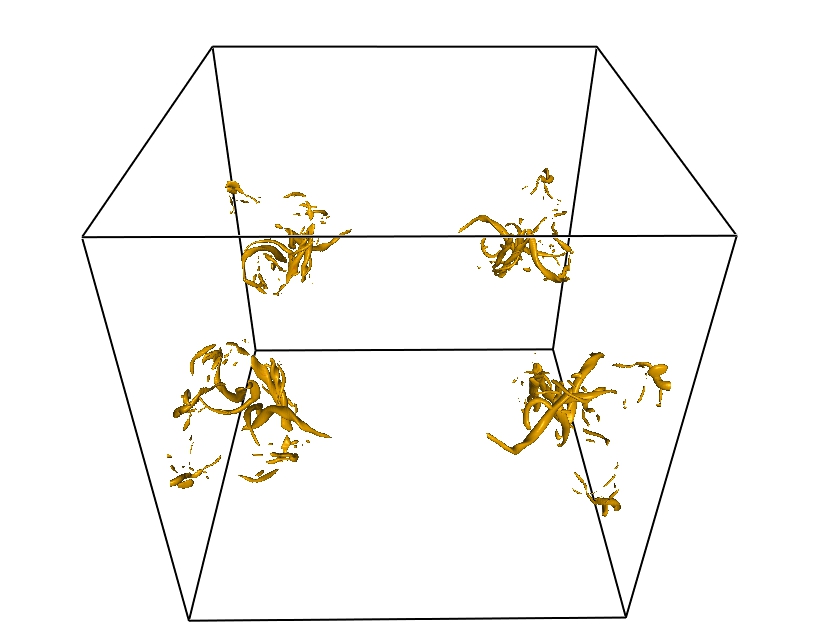}
c)\includegraphics[width=.45\linewidth]{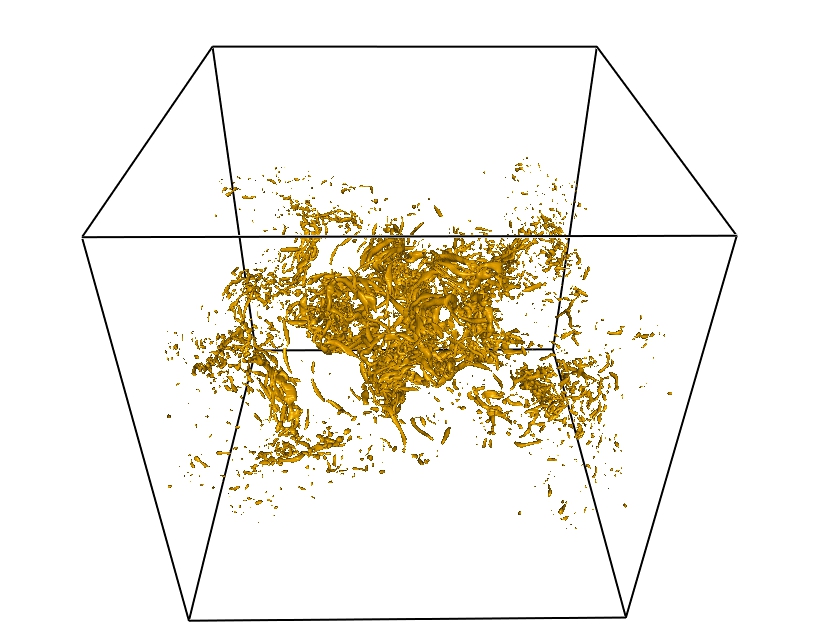}
d)\includegraphics[width=.45\linewidth]{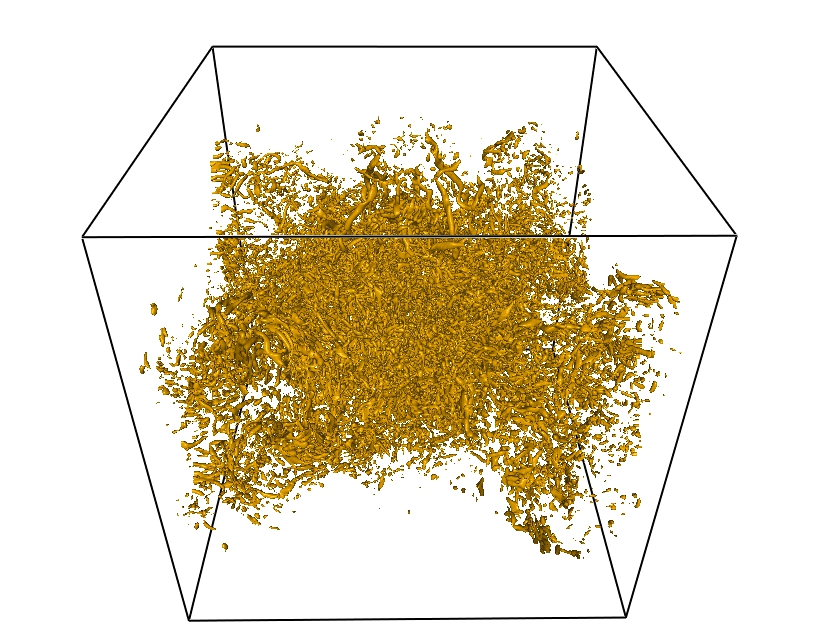}
}
   \caption{ Isosurfaces (at $12\%$ of maximum) of square vorticity $\omega^2=(\nabla \times {\bf v})^2$ at resolution $512^3$ in the impermeable box, $p=15$ and $t=4$, 6, 8, 10.}
 \label{fig:viz512_p_15}
\end{figure*}


\begin{figure*}
{\large
a)\includegraphics[width=.45\linewidth]{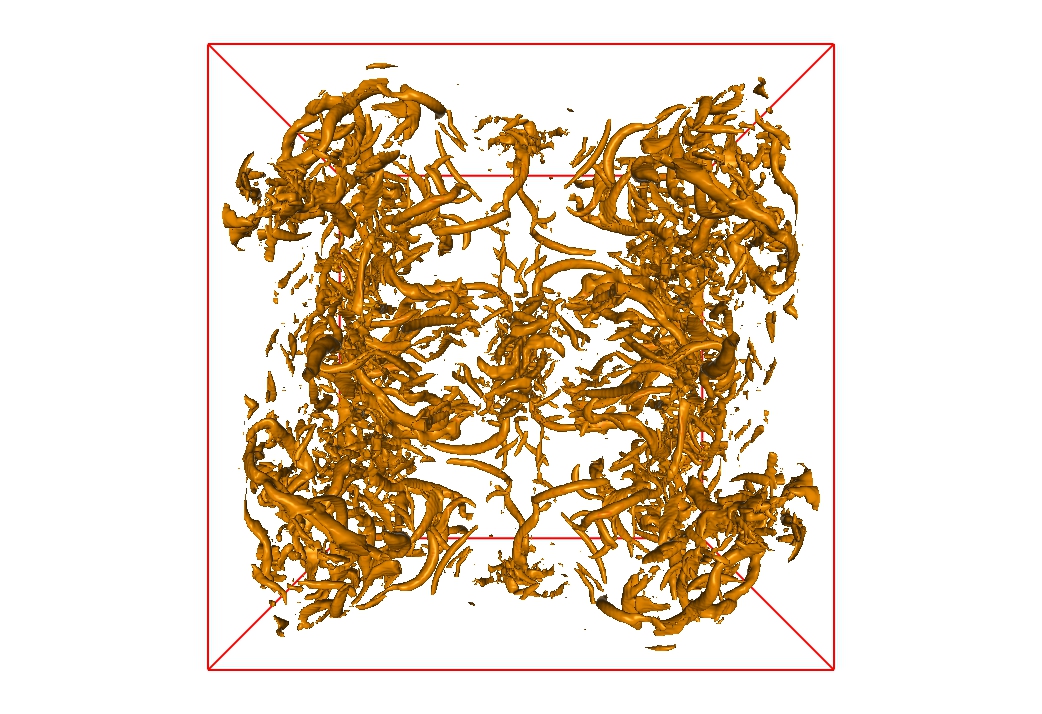}
b)\includegraphics[width=.45\linewidth]{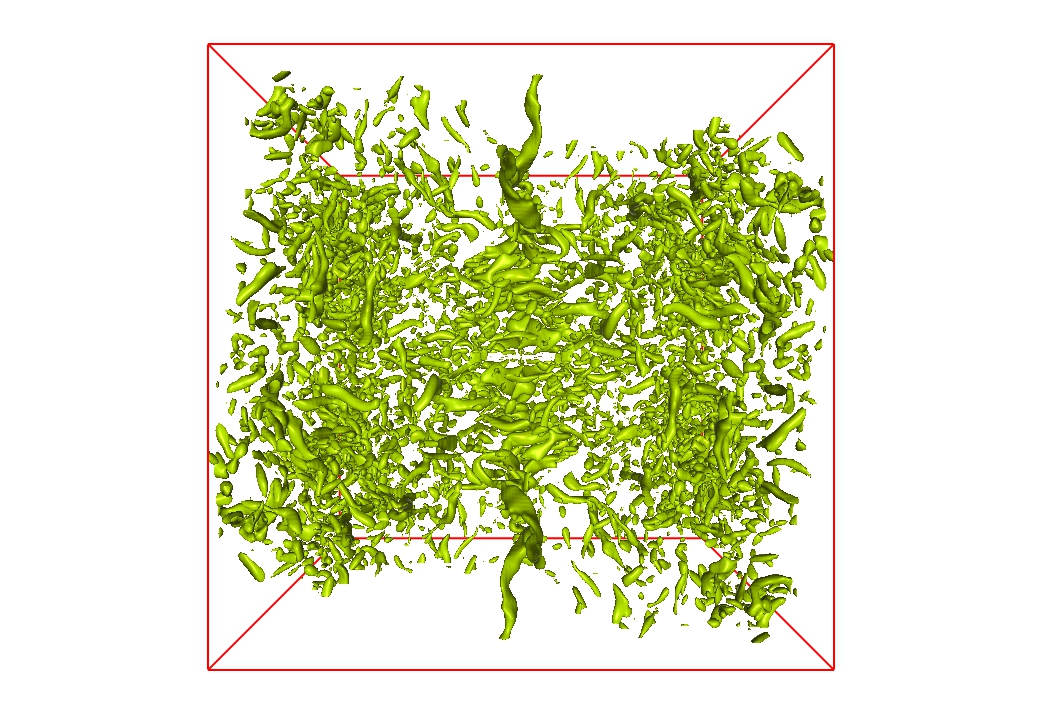}}
   \caption{Enlargement of a cubic subregion of width $\pi/3$ centered at $x=y=z=\pi/2$ of isosurfaces of square vorticity $\omega^2=(\nabla \times {\bf v})^2$ at resolution $1024^3$ and $t=9$. Left: $p=1$ and isosurface at $10\%$ of maximum; Right: $p=10$ and isosurface at $20\%$ of maximum.}
 \label{fig:viz1024_t_9}
\end{figure*}

We now examine the spatial structure of the flows computed with ordinary and with hyper viscosity.
We first recall that the mirror symmetries of the TG vortex with
respect to the planes $x=0,\pi$, $y=0,\pi$, and $z=0,\pi$ confine the flow inside the {\it impermeable box} formed by these planes.
The additional rotational symmetries of 
angle $\pi$ around the axes $x = z = \pi/2$, 
$y = z = \pi/2$, and $x = y = \pi/2$ are such that 
the early-time dynamics can be understood on and near the faces of the impermeable box. Computation of the flow shows that most of the dynamics of the flow for early-to-moderate times ($t < 4$) occurs near these faces. Simple dynamical considerations determine the behavior of the flow on the faces of the impermeable box, resulting in the rapid build-up of a vortex sheet. 
These considerations determine the dominant features of the flow correctly for times up to about $4$. 
A simple model of this phenomenon is given in appendix D of Brachet et al.~\cite{BRACHET:1983p4817}.
Because of the high Reynolds numbers considered in the present work, for all values of $p$ considered, the dynamics until $t=4$ is the same and essentially controlled by the inviscid dynamics. 

For the standard ($p=1$) NS equation, when $t$ increases beyond $4$, local vorticity maxima develop away from the walls of the impermeable box, suggesting the formation of new daughter vortices in the interior and a complicated flow structure, as displayed using VAPOR \cite{ncarVapor1}
in Fig. \ref{fig:viz512_p_1}. 
Similar visualizations obtained at $p=15$ are displayed in Fig. \ref{fig:viz512_p_15}. Comparing $p=1$ and $p=15$, we see that at $t=6$ the two visualizations are markedly different; the video in the supplemental material shows that the difference can be seen starting at around $t=5$. At later times, when thermalization occurs, the vortex tubes of Fig.~\ref{fig:viz512_p_1}d are dominated by fluctuations that take the form of {\it point blobs}, less elongated vorticity structures, as seen in Fig.~\ref{fig:viz512_p_15}d. 
This change of structure is best seen in Fig. \ref{fig:viz1024_t_9}, which shows visualizations of runs at resolution $1024^3$ at $t=9$  in a cubic subregion of width $\pi/3$ centered at $x=y=z=\pi/2$ for 
for $p=1$ (a) and $p=10$ (b). The vortex tubes that are clearly the dominant
structure in the $p=1$ case coexist in the $p=10$ case with point blobs that resemble the structures that appear in visualizations of the
truncated Euler equations in \cite{alexakis2019thermal}.



\subsection{ Thermalization and finite dissipation  } 

Looking at the previous sections we have two apparently contradictory results. On the one hand, based on figure \ref{fig:EnergyDiss} there is finite energy dissipation that is independent of the value of $p$ used. This indicates that this is an out-of-equilibrium dissipating system. 
On the other hand, the energy spectrum and structures  resemble those of the thermalized absolute equilibrium state \cite{kraichnan1973helical}. As discussed in the introduction, the thermalized state is realized in the truncated Euler equations, where a finite number of modes is kept and the energy is exactly conserved. Thus, if indeed there was a transition from the hyper-viscous Navier-Stokes equation to the truncated Euler equations, one would expect to see a suppression of the energy dissipation that is not observed here.

To resolve this discrepancy we
recall the arguments in \cite{frisch2008hyperviscosity}, in which 
the energy dissipation term in Fourier space
is written as 
\begin{equation}
    -\nu_p k^{2p} \hat{\bf u}_{\bf k} = \frac{U}{L} \left( \frac{k}{k_{_G}} \right)^{2p} \hat{\bf u}_{\bf k}
\end{equation}
where $\nu_p=U k_{_G}^{-2p}/L$ and $k_{_G}$ is a wavenumber in the dissipation range
chosen off-lattice so that no wavenumber is exactly equal to $k_{_G}$.
The limit $p\to \infty$ is then taken while keeping $k_{_G}$ fixed. This is similar
to the procedure followed here where we tuned $\nu_p$ so that the maximum of the
dissipation spectrum is smaller but close to the maximum wavenumber allowed by our grid.
In this case it is clear that, as $p\to \infty$, wavenumbers smaller than $k_{_G}$
will not feel the effect of viscosity while wavenumbers larger than $k_{G}$ will be 
suppressed. The whole system will thus resemble the truncated Euler system with 
$k_{G}$ acting as the truncation wavenumber. However for a finite $p$ (if the 
Fourier wavenumbers are sufficiently dense) there will be wavenumbers close to $k_{G}$
that will be effective at dissipating energy. 
To estimate the width of wavenumbers that effectively dissipate energy 
we consider the dissipation spectrum given by
$D(k)= \nu_p k^{2p} E(k)$ 
and assume that, at large $k$, the energy spectrum takes the form
$E(k) \propto k^{\alpha_p}e^{-k/k_d}$. Therefore 
$D(k) \propto k^{2p+\alpha_p}e^{-k/k_d}$, the integral of which 
gives the energy dissipation rate $\epsilon$. 
For large values of $p$, the dissipation spectrum is highly concentrated around 
a wavenumber $k_*$ and the integral $\int D(k)dk=\epsilon$ can be estimated
by writing
\begin{equation}
k^{2p+\alpha_p}e^{-k/k_d} = \exp\left[(2p+\alpha_p) \ln(k)-k/k_d \right]
\end{equation}
and using the steepest descent method. 
Expanding the argument of the exponential
around its maximum value at $k_*=(2p+\alpha_p) k_d$ 
with $D(k_*)=D_*=\nu_p(k_*/e)^{2p+\alpha_p}$, 
we see that the dissipation spectrum can be approximated as 
\begin{equation}
D(k) \simeq 
D_*
\exp\left[- \frac{1}{2} \left( \frac{q}{k_*/\sqrt{2p+\alpha_p}}\right)^2
\right] \end{equation}
where $q=k-k_*$, i.e. as a Gaussian centered at $k_*$ of width 
\begin{equation}\delta q \simeq \frac{k_*}{\sqrt{2p+\alpha_p }} \end{equation}
In our simulations, we tuned $\nu_p$ so that the maximum 
of the dissipation spectrum at $k_*$ remained fixed and close to the maximum wavenumber $k_{max}$. Therefore, the wavenumbers that dissipate effectively are those that 
are approximately within a distance of $\delta q$ away from $k_*$. 

For the dissipation to be suppressed, the width $\delta q$ 
must be smaller than the spacing $\delta k$ between two neighboring shells that contain at least one wavenumber 
in the discrete wavenumber space $\bf k \in \mathbb{N}^3$.
In three dimensions, the spacing between neighbouring spherical shells of radius $k$ is of the order
$\delta k \simeq 1/k$. (To illustrate this point, we note that the closest spherical shell to the shell containing the wavenumber ${\bf k}=(0,0,k)$ is that which contains the wavenumber ${\bf k^\prime}=(1,0,k)$ and has norm
$k^\prime=\sqrt{1+k^2}\simeq k + \frac{1}{2}k^{-1} + \dots$.) Equating $\delta q$ obtained from the steepest descent method with $\delta k=k^\prime-k$ we obtain that the system will behave like
the truncated Euler equations and energy dissipation will be suppressed when $\delta q \ll \delta k$ which implies that 
\begin{equation} 
p \gg k_*^4.
\end{equation}
Given that $k_*$ is our simulations is of the order $k_*\simeq 150 $ for the $512^3$
and $k_*\simeq 300 $ for the $1024^3$ simulation, it is clear why energy dissipation still persists in this system: the order $p$ would have to be around $10^{10}$ to see a strong suppression  of the energy dissipation. 
At $p$ of order $50$ and $100$, the system can still effectively dissipate energy since there are many wavenumbers inside the dissipating shell. 

What does change, however, as $p$ is increased is the 
number of triads $\bf k_1+k_2+k_3=0$ of interacting wavenumbers
that can transfer energy inside the dissipating shell.
This number is decreased drastically as $\delta q$ becomes smaller.
In contrast, the number of triads that redistribute energy among all wavenumbers and lead to the thermalized state remains fixed. 
The ratio of the two therefore becomes smaller as $p$ is increased (because $\delta q$ is decreased). This leads the system to a quasi-equilibrium state, in which the mean forward flux of energy (caused by the triads that transfer energy to the dissipating wavenumber shell) is subdominant to the fluctuations caused by the remaining triads that redistribute energy among modes leading to a thermalized state. This behavior has been observed recently by two of the authors \cite{alexakis2019energy}, who showed that the thermalized behavior can appear in forced and dissipated flows for the truncated Navier-Stokes system. It was shown that for a given injection rate as the viscosity is reduced, the system makes a transition to a quasi-equilibrium, with the appearance of a thermal spectrum, provided that $k_{max}\eta \ll 1 $. A similar situation occurs in the present system, not because of the reduction in viscosity, but because the dissipation is limited to wavenumbers inside a thin spherical shell. 

\section{Conclusion}

In this work we have examined decaying turbulence initiated by a Taylor-Green vortex using hyper-viscous numerical simulations for a wide range of orders of the hyper-viscous parameter $p$. We have shown that it is possible to integrate the Navier-Stokes equations with hyper-viscosity of order $p$ of 50 or 100, much higher than the values of 2 to 8 that have been previously studied
 \cite{borue1996numerical,borue1995forced,haugen2004inertial,lamorgese2005direct,spyksma2012quantifying,l1998universal}.
For all values of $p$ that we examined, the evolution of the total energy and its 
dissipation rate remained unaffected by the hyperviscosity and close to those of high-resolution $p=1$ runs.
 The spectra and the structures remain unaltered up to $t=5$, where almost inviscid dynamics are followed. At later times, however, the structures and the spectra diversify with the order of $p$. 
Even at these later times, the low-wavenumber portion of the spectra is surprisingly insensitive to the value of $p$ even for high values of $p$, but the inertial range (the part of the spectrum that displays a $k^{-5/3}$ scaling) does not extend to indefinitely high $k$.
As the order $p$ is increased,
a stronger bottleneck forms and the spatial  structures
change from vortex tubes to a mixture of vortex tubes and point blobs.
Further studies would shed light on the nature of high-$p$ hyper-viscosity, now that we have demonstrated its practical feasibility.

The main focus of this work was on the turbulent behavior at increasingly large values of $p$ and attempted to make the connection between the bottleneck effect that is present in the usual Navier-Stokes equation and the thermalization of the flow that develops in the truncated Euler equations. 
We showed that, as the hyper-viscosity order $p$ is increased, the energy spectrum approaches that of the thermalized state of the truncated Euler equations of Kraichnan \cite{kraichnan1973helical}. 
This transition was predicted in \cite{frisch2008hyperviscosity} by arguing that the hyper-viscous Navier-Stokes equations approach the truncated Euler equations as $p$
goes to infinity. 
This is the first time that this transition has been demonstrated in simulations of three-dimensional turbulence. 

Nonetheless for the hyper-viscous flow a finite dissipation rate independent of the value of $p$ persists, contrary to the situation for the truncated Euler system.
We argued that this behavior is due to the fact that for our grid resolutions and values of $p$, energy dissipation in Fourier space is concentrated in a spherical shell of width $\delta q \propto k_*/\sqrt{2p}$ that is thin but still much wider than the spacing between spherical wavenumber shells. Suppression of the energy dissipation was estimated to occur at much larger values of $p$.

In conclusion, a relationship has been established between the bottleneck in turbulence and flows in thermal equilibrium.  Our work has also demonstrated a continuous way to pass from the Kolmogorov spectrum to a thermalized spectrum. Unlike other systems that have demonstrated such a transition \cite{alexakis2019energy,shukla}, this path does not involve a discontinuous Galerkin truncation. 
 Future work could include the forced Taylor-Green vortex for which  
 long statistical averages can be performed.

\vspace{2pt}
\section{Acknowledgments}

The numerical simulations were performed using high performance computing resources provided by 
the Institut du Developpement
et des Ressources en Informatique Scientifique (IDRIS) of the Centre
National de la Recherche Scientifique (CNRS), coordinated by
GENCI (Grand Equipement National de Calcul Intensif) through grants A0050506421 and A0062A01119.
This work was also granted access to the HPC resources of MesoPSL financed by the Region 
Ile de France and the project Equip@Meso (reference ANR-10-EQPX-29-01) of the programme Investissements d'Avenir supervised by the Agence Nationale pour la Recherche. This work was also supported by the Agence Nationale pour la Recherche via ANR DYSTURB project No. ANR-17-CE30-0004.

\appendix
\section{Time-integration scheme}
\label{appendix}
 
We use an explicit second-order Runge-Kutta method to integrate the
nonlinear advective terms of the Navier-Stokes equations.
The linear viscous terms are often integrated via an implicit method,
in order to increase the timestep from that imposed by the viscous
stability requirement.  As the Reynolds number increases, this
constraint becomes less important than that imposed by the integration
of the advective terms, and hence explicit timestepping is sometimes
used for high Reynolds number simulations.  However, the use of
hyperviscosity poses a greater constraint on the timestep than
ordinary viscosity. An analogous problem occurs in integrating the
Kuramoto-Sivashinsky equation, which contains a fourth-order spatial
derivative and has led to the formulation of a modified exponential
scheme called the slaved scheme \cite{frisch1986viscoelastic}.  Here
we compare explicit timesetepping with exponential and modified
exponential timestepping for treating the hyperviscous terms.

To describe the exponential methods, we write the evolution equation for a mode
with wavenumber $k$ schematically as
\begin{equation}
  \partial_t u = - \nkp u + \mN(u)
\label{eq:PDE}\end{equation}
where $\mathcal{N}$ includes both the advective terms and the pressure
projection. Equation \eqref{eq:PDE} can be rewritten as an integral equation
\begin{equation}
u(t+\dt)=e^{-\nkp\dt}\left[u(t)+\int_t^{t+\dt}\hspace*{-0.3cm}d\tau\:e^{\nkp(\tau-t)}\mN(u(\tau))\right]
\label{eq:integ}\end{equation}
A first approximation, called time-splitting and used in both methods,
is $\mN(u(\tau)) \approx \mN(u_N(\tau))$,
where $u_{N}$ is the result of integrating $\partial_t u_N = \mN(u_N)$
from initial condition $u(t)$, in practice by the second-order Runge-Kutta scheme.

The exponential method further approximates the exponential $e^{\nkp(\tau-t)}$
in the integrand by 1, its value at the left endpoint, so that the integral becomes 
\begin{align}
\int_t^{t+\dt}\hspace*{-0.3cm}d\tau\:\mN(u_{N}(\tau)) \approx u_N(t+\dt)-u(t)
\end{align}
leading to the exponential scheme
\begin{align}
u(t+\dt)&\approx e^{-\nkp\dt}u_{N}(t+\dt)
\label{eq:exp}\end{align}

The modified exponential, or slaved, method approximates the exponential in
the integral by its average value
\begin{equation}
\frac{1}{\dt}\int_t^{t+\dt}\hspace*{-0.3cm}d\tau\:e^{\nkp(\tau-t)} = \frac{e^{\nkp\dt}-1}{\nkp\dt}  \end{equation}
leading to the slaved scheme
\begin{align}
  u(t+\dt)&\approx e^{-\nkp\dt}u(t) \nonumber\\
  &+ \frac{1-e^{-\nkp\dt}}{\nkp\dt}(u_N(t+\dt)-u(t))\label{eq:slaved}\\\nonumber
\end{align}
where the fraction in \eqref{eq:slaved} is evaluated as 1 if $\nkp\dt$
is less than $10^{-5}$.
Unlike the exponential scheme \eqref{eq:exp}, the
slaved scheme \eqref{eq:slaved} is exact for all $\dt$ in the special
case that $\mN(u)$ is a constant $\mN$:
\begin{align}
\intertext{Exact solution/Slaved method}
  u(t+\dt)&=e^{-\nkp\dt}u(t) + \frac{1-e^{-\nkp\dt}}{\dt\nkp}\dt\mN \\
  \intertext{Exponential method}
  u(t+\dt)&=e^{-\nkp\dt}\left[u(t) + \dt \mN\right]
\end{align}
For $\nkp\dt \rightarrow \infty$, the exact solution and
the result of the slaved method is $\mN/\nkp$,
while the solution obtained by the exponential method is zero.
A generalization of this argument shows that
if $\mN(u)$ varies slowly compared
with the timescale $1/(\nkp)$, then scheme \eqref{eq:slaved} is 
accurate even for large $\dt$.
Thus, this scheme yields correct results both for $\nkp\dt \ll 1$ and
$\nkp\dt \gg 1$ while making some error for $\nkp\dt\sim 1$.

\begin{figure}[!h]
\vspace*{0.5cm}
 \centerline{\includegraphics[width=\columnwidth]{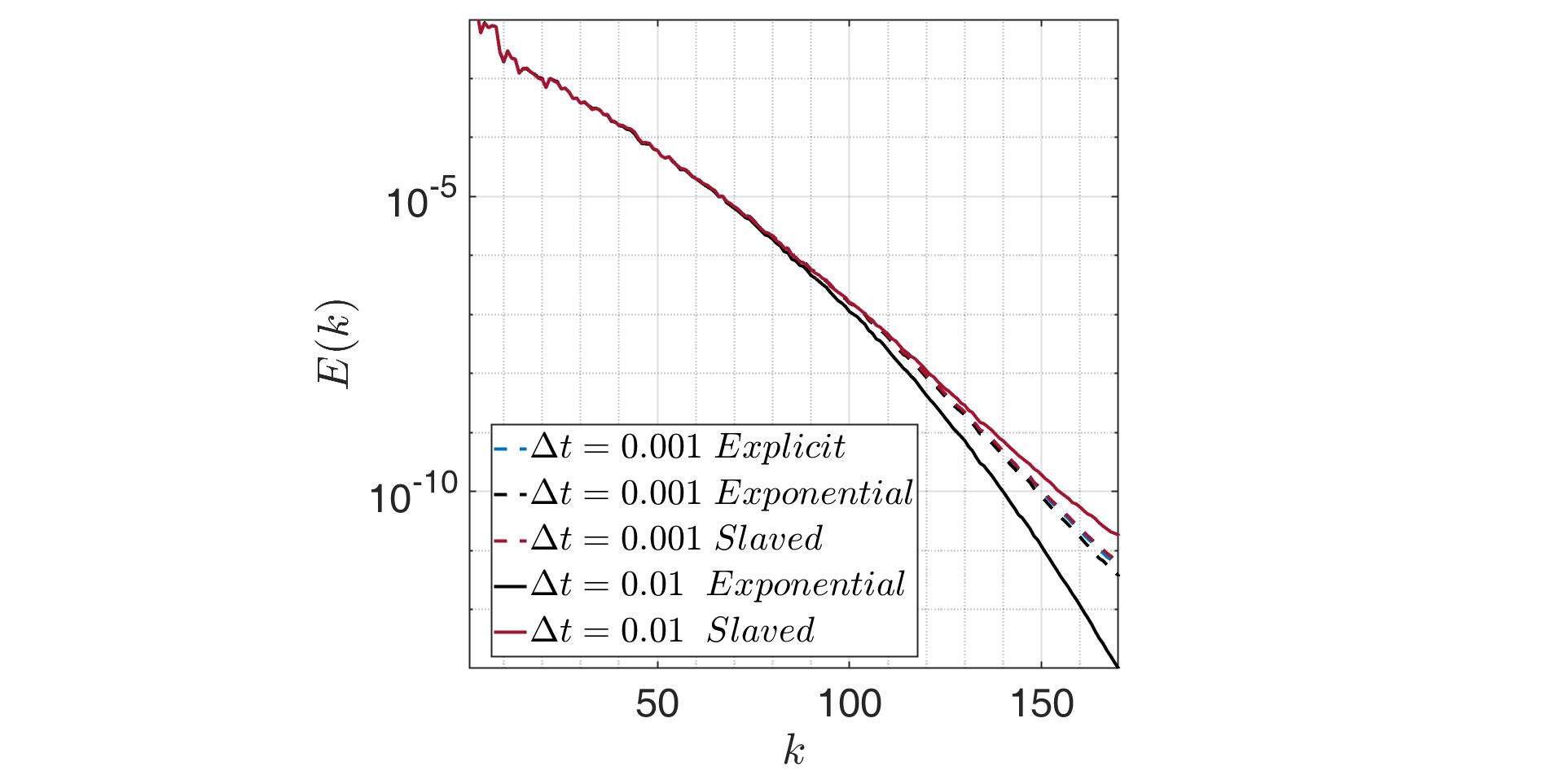}}
 \caption{Comparison of time-stepping schemes for $p = 2$, $N = 512$,
   $Re = 2.56\times 10^6$ for $\dt=0.001$ and $\dt=0.01$.  The
   nonlinear advective terms are integrated via the explicit
   second-order Runge Kutta scheme, while the linear hyperviscous
   terms are integrated using the explicit, exponential and slaved
   schemes. For $\dt=0.001$, the spectrum obtained using the three
   methods are quite close.  For $\dt=0.01$, the explicit scheme
   diverges, while the slaved scheme is more accurate than the
   exponential scheme, i.e.~the spectrum more closely resembles that
   obtained for $\dt=0.001$.}
 \label{fig:numerical_comparison}
 \end{figure}

Figure \ref{fig:numerical_comparison} compares the 
energy spectrum that results from using these three schemes for
a case with mild hyperviscosity $p=2$, spatial resolution $N=512$,
Reynolds number $Re=2.56\times 10^6$, for two times, $\Delta t=0.001$
and $\Delta t = 0.01$. For $\Delta t=0.001$, all three schemes
give satisfactory results. For $\Delta t=0.01$, the explicit scheme
diverges, while the slaved scheme yields a spectrum that is
more accurate than the exponential scheme, 
i.e.~the spectrum is closer to that obtained for $\dt=0.001$.

\newpage
\bibliography{bibli}
\end{document}